\documentclass[aps]{revtex4}
\usepackage{epsfig, amssymb, latexsym, amsfonts, amsmath, amsthm}

\begin{document}
\sloppy

\title{Twist cut-off for scalar off-cone operators in $q$-space}

\author{ J\"org Eilers }
\email{ eilers@itp.uni-leipzig.de }
\affiliation{ Center for Theoretical Studies and Institute of Theoretical Physics, Leipzig University, Augustusplatz~10, D-04109~Leipzig,
Germany}

\author{ Bodo Geyer }
\email{ geyer@itp.uni-leipzig.de }
\affiliation{ Center for Theoretical Studies and Institute of Theoretical Physics, Leipzig University, Augustusplatz~10, D-04109~Leipzig,
Germany}

\date{\today}

\begin{abstract}
\vspace*{0.1cm} \noindent
We show that the twist decomposition of non-local scalar off-cone QCD-operators, which in $x-$space is infinite, after Fourier
transformation with the singular functions $1/(x^2 - {\text{i}} \epsilon)^\lambda$ for $\lambda = 1$ and $2$ terminates at leading and
sub-leading twist, respectively. Therefore, only a finite number of twists contribute to the target mass corrections of light-cone dominated
scattering processes. Their explicit expressions may be read off directly from the twist decomposition in $x-$space. \\

\vspace*{0.0cm}

\end{abstract}
\maketitle
\section{Introduction}
\noindent
Non-perturbative parton distribution amplitudes parameterizing the matrix elements of non-local quark-antiquark operators play a central
role in phenomenological considerations of hard hadronic processes. Thereby, different processes are governed by one and the same (set of)
nonlocal operators which occur in a quantum field theoretic description via the non-local light-cone expansion \cite{AZ78,ZAV,MRGHD}. For
deep inelastic lepton-hadron scattering and Drell-Yan processes the parton distributions are given as forward matrix elements of bilocal
light-ray operators. For (deeply) virtual Compton scattering and hadron wave functions the so-called double distributions and hadron
distribution amplitudes, respectively, are given by corresponding non-forward matrix elements.

The dependence of these distribution amplitudes on the momentum transfer $Q^2$ may be determined perturbatively. In higher orders, besides
radiation corrections, there are various concurring effects due to the (higher) twists of operators appearing in the light-cone expansion as
well as due to target mass contributions being related to the traces of the off-cone operators of definite twist. Here, the notion of {\em
geometric twist} $\tau$ = (canonical) dimension $d$ - spin $j$ is used as has been introduced by Gross and Treiman \cite{GT71} for local
operators as a purely group theoretical concept.

Using a purely group theoretical procedure the (finite) twist decomposition of non-local light-ray operators in configuration space, as
far as they are relevant for light-cone dominated hadronic processes, has been performed in Ref.~\cite{GLR99,GLR00}. However, if one wants
to calculate target-mass corrections for scattering processes one is forced to consider the (infinite) twist decomposition off the
light-cone thereby taking into account all trace terms which, after Fourier transformation into momentum space, lead to contributions
suppressed by powers of $M^2/Q^2$. However, a {\em complete} off-cone twist-decomposition of non-local operators is only possible when their
$n$th moments are totally symmetric, i.e., have symmetry type $\left[ n \right]$. In general, non-local operators occur which are related to
irreducible tensor representations of the Lorentz group whose symmetry type is determined by non-trivial Young patterns~\cite{GLR00}. The
twist decomposition in $x-$space has been determined completely for the totally symmetric case as well as partially for the physically
relevant non-trivial QCD operators \cite{GLR01}. However, that consideration missed the final Fourier transformation. This will be
considered here in the most simplest case of totally symmetric scalar operators.

Target mass corrections in unpolarized deep inelastic scattering has been treated group theoretically for the first time by Nachtmann
\cite{Nachtmann73} by expanding into a series of Gegenbauer polynomials. The same procedure was then applied later on by various authors to
calculate target mass corrections in polarized deep inelastic scattering \cite{BE76,BE76_b,W77,MU80,KU95} and to determine meson mass
corrections \cite{B99,BB99}. Ref.~\cite{BM01} treats twist--2 target mass corrections in deeply virtual Compton scattering. -- Another
method for the construction of mass corrections in the unpolarized forward case was given by Georgi and Politzer~\cite{GP}. This method has
been applied to polarized forward scattering in Ref.~\cite{PR98,BT99}.

Let us begin with the expression for the amplitude of virtual Compton scattering,
\begin{equation}
\label{CA_nonf}
 T_{\mu\nu}(P_i,Q; S_i) = {\text{i}} \int {\mathrm{d}}^4 \! x \; {\text{e}}^{{\text{i}} qx} \bigl< P_2,S_2 \bigl|\; RT \left( J_\mu(x/2)
J_\nu(-x/2) \, \mathcal{S} \right) \, \bigr| P_1,S_1 \bigr> \; ,
\end{equation}
where $P_1 (P_2)$ and $S_1 (S_2)$ are the momenta and spins of the incoming (outgoing) hadrons, $Q^2=-q^2,\,q= q_2-q_1= P_1-P_2$ denotes
the momentum transfer and $\cal S$ is the (renormalized) $S-$matrix. The renormalized time-ordered product of the two electro-magnetic
currents in $J_\mu$ when approximating the quark propagator near the light-cone by its most singular parts is given by the well-known
expression ($\kappa = 1/2$):
\begin{eqnarray}
\label{str_wick}
\nonumber
 T\left[ J_\mu\left( \kappa x \right)J_\nu\left( -\kappa x \right) \right] \,
&\approx&
 \left[ \frac{1}{2\pi^2\left( x^2 - \text{i} \epsilon \right)^2} + \frac{m^2}{8\pi^2\left( x^2 - \text{i} \epsilon \right) } \right]
\\
\nonumber
&&
 \qquad \times \Bigl( g_{\mu\nu} \, O\left( \kappa x, - \kappa x \right) - 2 \, x_{\left\{ \mu \right.} O_{\left.
\nu \right\}} \left( \kappa x, - \kappa x \right) - {\text{i}} \,{\epsilon_{\mu\nu}}^{\alpha\beta} x_\alpha
O_\beta^5\left( \kappa x, - \kappa x \right) \Bigr)
\\
&&
 - \; \frac{ { \text{i} } \, m }{4\pi^2\left( x^2 - {\text{i}} \epsilon \right) } \;\, \biggl( g_{\mu\nu} \, N\left( \kappa x, - \kappa x
\right) + M_{\left[ \mu\nu \right]} \left( \kappa x, - \kappa x \right) \biggr) \, .
\end{eqnarray}
Here, the 'centered' non-local chiral-even (axial) vector operators $O_\mu^{(5)}\left( \kappa x, - \kappa x \right)$ and the scalar
operator $O\left( \kappa x, - \kappa x \right) = x^\mu O_\mu\left( \kappa x, - \kappa x \right) $ are given by the following
(anti)symmetrized operators (whose leading twist is $\tau_0$ is 2 and 3)
\begin{eqnarray}
\label{vector_op}
\nonumber
 O_\mu\left( \kappa x, - \kappa x \right)
&=&
 :\!\bar\psi\left( \kappa x \right)\,\gamma_\mu \,\psi\left( -\kappa x \right)\!: - :\!\bar\psi\left( -\kappa x \right)\, \gamma_\mu
\,\psi\left( \kappa x \right)\!: \;,
\\
\nonumber
 O_\mu^5\left( \kappa x, - \kappa x \right)
&=&
 :\!\bar\psi\left( \kappa x \right)\, \gamma^5 \gamma_\mu \,\psi\left( -\kappa x \right)\!: + :\!\bar\psi\left( -\kappa x \right)\, \gamma^5
\gamma_\mu \,\psi\left( \kappa x \right)\!: \;,
\\
 O\left( \kappa x, - \kappa x \right)
&=&
 :\!\bar\psi\left( \kappa x \right)\, x \!\!\! /\, \,\psi\left( -\kappa x \right)\!: - :\!\bar\psi\left( -\kappa x \right)\,x \!\!\! /\,
\,\psi\left( \kappa x \right)\!: \; ,
\end{eqnarray}
whereas the 'centered' chiral-odd skew tensor operator $M_{\left[ \mu\nu \right]} \left( \kappa x, - \kappa x \right)$ and the scalar
operator $N\left( \kappa x, - \kappa x \right)$ are given by the following (anti)symmetrized operators (whose leading twist is $\tau_0=3$)
\begin{eqnarray}
\nonumber
 M_{\left[ \mu\nu \right]} \left( \kappa x, - \kappa x \right)
&=&
 :\!\bar\psi\left( \kappa x \right) \, \sigma_{\mu\nu} \, \psi\left( -\kappa x \right)\! : - :\!\bar\psi\left( -\kappa x \right) \,
\sigma_{\mu\nu} \, \psi\left( \kappa x \right)\! :\;,
\\
 N\left( \kappa x, - \kappa x \right)
&=&
 :\!\bar\psi\left( \kappa x \right) \; \psi\left( -\kappa x \right)\! : + \, :\!\bar\psi\left( -\kappa x \right) \, \psi\left( \kappa x
\right)\!: \; .
\end{eqnarray}

In the following we restrict our considerations to the operator content of the Compton amplitude. Namely, we study twist contribution of
the scalar operators $O$ and $N$ to the Compton amplitude (\ref{CA_nonf}) and, for completeness, also the scalar operator
$M\left( \kappa x, - \kappa x \right)$ which is constructed out of the skew tensor operator
$M_{\left[ \mu\nu \right]} \left( \kappa x, - \kappa x \right)$ according to
\begin{equation*}
 M\left( \kappa x, - \kappa x \right) = \partial_x^\mu \, x^\nu \, M_{\left[ \mu\nu \right]} \left( \kappa x, - \kappa x \right) \; .
\end{equation*}
Obviously, the operator $ O \left( \kappa x , - \kappa x \right) $ contributes to the leading light-cone singularity being independent of
the quark-mass $m$ and also to the lower light-cone singularities being proportional to $m^2$. The operator
$N\left( \kappa x, - \kappa x \right)$ only contributes to quark-mass terms and therefore is not present if only leading contributions are
considered.

\section{complete twist decomposition of scalar off-cone operators in $x$-space}

Let us briefly repeat the procedure and those results of Ref.~\cite{GLR01} which are relevant here, thereby also introducing the
terminology. First, one performs -- without taking into account the 'external' contractions with $x^\nu$ or $\partial_x^\mu x^\nu$ -- a
formal Fourier transformation of the non-local operators followed by an expansion into local operators,
\begin{equation}
\label{formal_fourier}
 \mathcal{O}_{\Gamma (x)}\left( \kappa x, - \kappa x \right) = \int {\mathrm{d}}^4 \! u \; \mathcal{O}_{\Gamma(x)}\left( u \right) \,
{\text{e}}^{-{\text{i}}\kappa \, ux} = \sum_{n=0}^\infty \frac{\left( -{\text{i}}\kappa \right)^n}{n!} \mathcal{O}_{\Gamma n+i}\left( x
\right) \; .
\end{equation}
Here, $\mathcal{O}$ generically denotes $\{N, O, M\}$ and $\Gamma(x)$ denotes
$\left\{ {\mathbb{I}}, x \!\!\! /\, , \partial_x^\rho \, x^\sigma \sigma_{\rho\sigma} \right\}$
with $i =\left\{ 0, 1 , 0 \right\}$ counting powers of the external variable $x$.
Furthermore, we treat the unsymmetrized operators only, but the final (anti)symmetrized expressions can easily be obtained due to the
$\kappa-$dependence. For notational simplicity we understand the factor $1/(2\pi)^4$ to be included into the measure of the Fourier
transformation (\ref{formal_fourier}). (It is important to note that this Fourier transformation has nothing to do with the (inverse)
Fourier transformation w.r.t. $q$ of the Compton amplitude (\ref{CA_nonf})). The explicit expressions for the moments
$\mathcal{O}_{\Gamma n+i}\left( x \right)$ introduced above read
\begin{alignat}{2}
\label{moment_N}
 N_n\left( x \right)
&=
 \int {\mathrm{d}}^4 \! u \; N\!\left( u \right) \, \left( ux \right)^n &
&= \int {\mathrm{d}}^4 \! u \;
 \left( \bar \psi \psi \right)\! \left( u \right) \, \left( ux \right)^n \;,
\\
\label{moment_O}
 O_{n+1}\left( x \right)
&=
 \int {\mathrm{d}}^4 \! u \; O^\rho\!\left( u \right) \,x_\rho \left( ux \right)^n &
 &= \hbox{\Large$\frac{1}{n+1}$}\int {\mathrm{d}}^4 \! u \; \left( \bar \psi \gamma^\rho \psi \right)\! \left( u \right) \,
 \partial_\rho^u \left( ux \right)^{n+1}\;,
\\
\label{moment_M}
 M_n\left( x \right)
&=
 \int {\mathrm{d}}^4 \! u \; M^{\left[ \rho\sigma \right]}\! \left( u \right) \,\partial^x_\rho \, x_\sigma \,
 \left( ux \right)^n &
 &= \int {\mathrm{d}}^4 \! u \; \left( \bar \psi \, \sigma^{\rho\sigma} \psi \right)\!\left( u \right)\,
 \partial^u_\sigma \, u_\rho \left( ux \right)^n \; .
\end{alignat}

Restricting onto the light-cone, $x \rightarrow \tilde x, {\tilde x}^2 =0$, the twist decomposition is finite for all non-local
QCD-operators \cite{GLR99}. Especially, for scalar operators only the leading twist occurs, i.e.,
$ \mathcal{O}_{\Gamma (\tilde x)}\left( \kappa\tilde x, - \kappa\tilde x \right) = \mathcal{O}^{{\text{tw}} \tau_0}_{\Gamma (\tilde x)}\left( \kappa\tilde x, - \kappa\tilde x \right) $
with $\tau_0 = \{3, 2,3\}$, respectively, while higher twists occur for vector and tensor
operators.

However, off-cone the twist decomposition is infinite -- also for scalar operators. In Ref.~\cite{GLR01} using the group theoretical
method of Bargmann and Todorov \cite{BT77} the twist decomposition has been given as follows:
\begin{eqnarray}
\label{O_op}
 \mathcal{O}_{\Gamma n+i}\left( x \right) &=& \sum_{j=0}^{\left[ \frac{n+i}{2} \right]} \frac{ \left( n+i+1-2j \right)! }{ 4^j \, j! \,
\left( n+i+1-j \right)! } \left( x^2 \right)^j \mathcal{O}_{\Gamma\ n+i-2j}^{ {\text{tw}}\left( \tau_0+2j \right) }\left( x \right)\;.
\end{eqnarray}
In Ref.~\cite{GLR01} erroneously this summation over operators of definite twist was written with an additional factor
$\left( -1 \right)^j$. Here, the local scalar operators of definite twist are given by
\begin{eqnarray*}
 \mathcal{O}_{\Gamma\ n+i-2j}^{ {\text{tw}}\left( \tau_0+2j \right) }\left( x \right) &=& H_{n+i-2j} \left( x^2, \Box_x \right) \; \Box_x^j
\; O_{\Gamma n+i}\left( x \right)\;,
\end{eqnarray*}
where the operator $ H_n\left( x^2, \Box_x \right) $ which projects onto (traceless) scalar harmonic tensors is defined as
\begin{equation}
\label{def_Hn}
 H_n\left( x^2, \Box_x \right) = \sum_{k=0}^{\left[ \frac{n}{2} \right]} \frac{ \left( -1 \right)^k \, \left( n-k \right)! }{ 4^k \, k! \,
n! } \left( x^2 \right)^k \Box_x^k \; .
\end{equation}
It is obvious from expression (\ref{O_op}) that the twist decomposition of the operators listed above is finite in $x$-space iff the
operators are taken on-cone since each contribution beyond leading twist is multiplied by an additional factor of $x^2$ which vanishes in
the limit $x^2 \rightarrow 0$.

The local operators of lowest twist which are relevant in the following read \cite{GLR01}:
\begin{eqnarray}
\label{N3n}
 N^{\text{tw3}}_n\left( x \right)
&=&
 \int {\mathrm{d}}^4 \! u \; \left( \bar \psi \psi \right)\! \left( u \right) \, h^1_n\left( u,x \right) \;,
\\
\label{M3n}
 M^{\text{tw3}}_n\left( x \right)
&=&
 \int {\mathrm{d}}^4 \! u \; \left( \bar \psi \, \sigma^{\rho\sigma} \psi \right)\!\left( u \right)\, \,u_\rho x_\sigma \, h^2_{n-1} \left(
u,x \right) \;,
\\
\label{O2n}
 O^{\text{tw2}}_{n+1}\left( x \right)
&=&
 \hbox{\large$\frac{1}{n+1}$}\int {\mathrm{d}}^4 \! u \; \left( \bar \psi \gamma^\rho \psi \right)\! \left( u \right) \, \left( x_\rho
h^2_{n}\left( u,x \right) - \hbox{\Large$\frac{1}{2}$} u_\rho x^2 h^2_{n-1}\left( u,x \right) \right)\;,
\\
\label{O4n}
 O^{\text{tw4}}_{n-1}\left( x \right)
&=&
 n \int {\mathrm{d}}^4 \! u \; \left( \bar \psi \gamma^\rho \psi \right)\! \left( u \right) \, \left( x_\rho u^2 h^2_{n-2}\left( u,x \right)
- \hbox{\large$\frac{1}{2}$} u_\rho u^2 x^2 h^2_{n-3}\left( u,x \right) +2 u_\rho h^1_{n-1}\left( u,x \right) \right)\;,
\end{eqnarray}
with
\begin{alignat*}{4}
h_n^\nu\left( u,x \right) &:\;= \left( \hbox{\large$\frac{1}{2}$}\sqrt{u^2
x^2} \right)^n \; C_n^{\nu}\left( \frac{ux}{\sqrt{x^2 u^2}} \right) & \quad
\text{for} \quad & n \geq 0 \quad \text{and} \quad \left( \nu,n \right) \,
\neq \left( 0,0 \right)\;,
\\
h_n^\nu\left( u,x \right) & :\;= 0 & \quad\text{for}\quad & n < 0\;,
\end{alignat*}
where $C^\nu_n\left( z \right)$ are the Gegenbauer polynomials (see Ref.~\cite{PBM}, Appendix II.11).

\section{complete twist decomposition of scalar off-cone operators in $q$-space}
In this section it will be proven by explicit construction that the infinite off-cone twist decomposition becomes finite after Fourier
transformation with appropriate singular functions $1/\left( x^2-{\text{i}}\epsilon \right)$ and $1/\left( x^2-{\text{i}}\epsilon \right)^2$
appearing in the expansion (\ref{str_wick}). Depending on the order of the light-cone singularity the switch to $q$-space projects out
leading and also sub-leading twist, respectively, of the operator under consideration. (Since $M\left( \kappa x , - \kappa x \right)$ is
constructed out of $M_{\left[ \mu\nu \right]}$ we will treat it with the same pre-factor.)

\subsection{ Scalar operators multiplied by the sub-leading light-cone singularity $ 1 / (x^2 - {\text{i}} \epsilon) $ }
Let us begin with the Fourier transformation of
$N^{\text{tw3}} \left( \kappa x , - \kappa x \right) / \left( x^2 - {\text{i}} \epsilon \right) $ as the simplest example:
\begin{eqnarray}
\nonumber
 \int \frac{{\mathrm{d}}^4 \! x}{2\pi^2} \; {\text{e}}^{{\text{i}} qx} \frac{1}{ x^2 - {\text{i}}\epsilon} \; N^{\text{tw3}}\left( \kappa x,
- \kappa x \right)
& = &
 \sum_{n=0}^\infty \frac{\left( -{\text{i}}\kappa \right)^n}{n!} \int \frac{{\mathrm{d}}^4 \! x}{2\pi^2} \; {\text{e}}^{{\text{i}} qx}
\frac{1}{x^2 - {\text{i}}\epsilon} H_{n}\left( x^2,\Box_x \right) \int {\mathrm{d}}^4 \! u \; N\! \left( u \right) \,\left( ux \right)^n
\\
\label{H_n}
& = &
 - {\text{i}} \sum_{n=0}^\infty \kappa^n \int {\mathrm{d}}^4 \! u \; N\! \left( u \right) \; H_{n}\left( u^2,\Box_u \right) \;
\mathbf{h}_n^1(u,q)
\\
\label{ergebnis_local}
& = &
 - {\text{i}} \sum_{n=0}^\infty \kappa^n \int {\mathrm{d}}^4 \! u \; N\! \left( u \right) \; \mathbf{h}_n^1(u,q)\;.
\end{eqnarray}
Here we have used the fact that, since the monomial $\left( ux \right)^n$ is symmetric in both variables,
$H_n\left( x^2,\Box_x \right)\left( ux \right)^n = H_n\left( u^2,\Box_u \right)\left( ux \right)^n$ and that the Fourier transformation of
$\left( ux \right)^n/\left( x^2 - {\text{i}} \epsilon \right)$ may be written in terms of Gegenbauer polynomials, namely, it holds
\begin{equation}
\label{formel_wichtig}
 \int \frac{{\mathrm{d}}^4 \! x}{2\pi^2} \; {\text{e}}^{{\text{i}} qx} \frac{ \left( ux \right)^n }{ x^2 - {\text{i}}\epsilon} = -
{\text{i}}^{n+1} \; n! \; \mathbf{h}^1_n\left( u,q \right)\;,
\end{equation}
with $\mathbf{h}^\nu_n(u,q)$ defined by
\begin{eqnarray}
 \mathbf{h}_n^\nu\left( u,q \right) & :=& \frac{2^{n+\nu}\,\Gamma(\nu)}{(q^2+{\text{i}} \varepsilon)^{n+\nu}} \; h^\nu_n\left( u,q
\right)\;.
\end{eqnarray}
Formula (\ref{formel_wichtig}) can be derived by representing the monomial $\left( ux \right)^n $ by $ (-{\text{i}} u \partial_q)^n$
outside the integral and using the standard result that the Fourier transformation of $1/(x^2 - {\text{i}} \epsilon)$ is given by
$ - 2 {\text{i}} / \left( q^2 + {\text{i}} \epsilon \right) $. Performing all subsequent $q-$differentiations one obtains the desired
result.

Just like $h^1_n(u,x)$ is harmonic in $u$ and $x$, the polynomial $\mathbf{h}^1_n(u,q)$ is harmonic in $u$ and $q$ and therefore the
projection operator $H_n\left( u^2,\Box_u \right)$ in (\ref{H_n}) reduces to the identity, giving immediately the local version
(\ref{ergebnis_local}). One may reformulate this result by saying that the trace terms subtracted by $H_n\left( x^2,\Box_x \right)$ in
$x$-space (cf. Eq.~(\ref{def_Hn})) are cut-off by Fourier transformation into the $q$-space.

In fact, the above result coincides with the Fourier transformation of the undecomposed operator. Namely,
\begin{eqnarray*}
&&
\int \frac{{\mathrm{d}}^4 \! x}{2\pi^2} {\text{e}}^{{\text{i}} qx} \frac{1}{ x^2 - {\text{i}}\epsilon} \; \left( N\left( \kappa x, - \kappa
x \right) - N^{\mathrm tw 3} \left( \kappa x, - \kappa x \right) \right)
\\
&&
\qquad\qquad\qquad = -{\text{i}} \sum_{n=0}^\infty \kappa^n \sum_{j=1}^{\left[ \frac{n}{2} \right]} \frac{ \left( n+1-2j \right)! }{ 4^j \,
j! \, \left( n+1-j \right)! } \int {\mathrm{d}}^4 \! u \; N \left( u \right) \;(u^2)^j \, \Box_q^j \; H_{n-2j}(u^2, \Box_u) \;
\mathbf{h}_{n-2j}^1\left( u,q \right)\;
\end{eqnarray*}
vanishes term by term since $\Box_q^j \; \mathbf{h}_{n-2j}^1\left( u,q \right) =0$ for $j\geq 1$. With this result in mind one easily
convinces oneself that the Fourier transform of the function
\begin{equation}
{h}_{n}^1\left( u,x \right)= \sum_{k=0}^{\left[ \frac{n}{2} \right]} \frac{(-1)^k \left( n-k \right)!} { 4^k \, k! \, \left( n-2k \right)! }
\left( ux \right)^{n-2k}\left( u^2 x^2 \right)^k
\end{equation}
multiplied by $1/\left( x^2-{\text{i}}\epsilon \right)$ coincides with the Fourier transform of
$(ux)^n/\left( x^2-{\text{i}}\epsilon \right)$ given by Eq.~(\ref{formel_wichtig}). In fact, the Fourier transform of all functions
$h^\nu_n\left( u,x \right)/\left( x^2-{\text{i}}\epsilon \right)$ is proportional to $\mathbf{h}^1_n\left( u,q \right)$.

Comparing the local operators of Eq.~(\ref{ergebnis_local}) with with those of Eq.~(\ref{N3n}) one observes that the structure in terms of
Gegenbauer polynomials is one and the same in $x$- and in $q$-space. Of course, in case of the scalar operator $N$ this result is not new
and has first been given by Nachtmann in Ref.~\cite{Nachtmann73} in the case of forward scattering.

Finally, performing in expression (\ref{ergebnis_local}) the summation over $n$ in order to get a non-local operator, one may use the
summation properties of Gegenbauer polynomials (cf., Ref.~\cite{PBM}, Eq. 5.13.1.1) leading to
\begin{equation}
\label{summe}
 \sum_{n=0}^\infty \kappa^n \, \mathbf{h}_n^\nu(u,q) = 2^\nu \, \Gamma\left( \nu \right)\,\left( (q-\kappa u + {\text{i}}\epsilon)^{2}
\right)^{-\nu} \; ,
\end{equation}
with the result
\begin{eqnarray}
\nonumber
 \int \frac{{\mathrm{d}}^4 \! x}{2\pi^2} \; {\text{e}}^{{\text{i}} qx} \frac{1}{ x^2 - {\text{i}}\epsilon} \; N^{\text{tw3}}\left( \kappa x,
- \kappa x \right) &=& \frac{-2\,{\text{i}}}{q^2+{\text{i}}\epsilon} \int {\mathrm{d}}^4 \! u \; N\! \left( u \right) \frac{1}{{\left( 1 - 2
\,\kappa \, X + \kappa^2 \, X^2 M^2 \right) }}\;.
\end{eqnarray}

The variables $X = \left( uq \right) / q^2 $ and $M^2 = u^2 q^2 / \left( uq \right)^2$ are dimensionless fractions, which already have
been introduced in \cite{BM01} where they are called $-1/\Xi$ and $-\mathcal{M}^2$, respectively. If matrix elements are taken the variable
$u$ gets some combination of the momenta. Then, in case of forward scattering $X$ becomes the inverse Bjorken variable $x_B$ and in more
general processes \cite{BEGR02} it generates all scaling variables necessary to fix the kinematic domain of the process, e.g., $\xi$ and
$\eta$ in the case of non-forward scattering. Analogously, $M^2$ picks up all the products between incoming and outgoing hadronic momenta
and therefore generates all target mass corrections (see Ref.~\cite{GLR01}, formulae (2.11) to (2.14), for details of the parameterization
of hadronic matrix elements in different scattering processes).

Equipped with these basic tools we can immediately write down the Fourier transformation of the remaining scalar non-local operators when
multiplied with the pre-factor $ 1/\left( x^2 - {\text{i}} \epsilon \right)$. For the operator
$M^{\text{tw3}} \left( \kappa x , - \kappa x \right)$ we get
\begin{eqnarray}
\label{M_fourier}
 \int \frac{{\mathrm{d}}^4 \! x}{2\pi^2} \; {\text{e}}^{{\text{i}} qx} \frac{1}{ x^2 - {\text{i}}\epsilon} \; M^{\text{tw3}}\left( \kappa x,
- \kappa x \right)
&=&
 - {\text{i}} \sum_{n=0}^\infty \kappa^n \int {\mathrm{d}}^4 \! u \; M^{ \left[ \rho\sigma \right] } \left( u \right) \; \partial_\sigma^u
\, u_\rho \; \mathbf{h}_n^1(u,q)
\\
\label{M_fourier_b}
&=&
 - {\text{i}} \sum_{n=0}^\infty \kappa^n \int {\mathrm{d}}^4 \! u \; M^{ \left[ \rho\sigma \right] } \left( u \right) \; q_\sigma \, u_\rho
\, \mathbf{h}_{n-1}^2(u,q)
\\
\nonumber
&=&
 \frac{4 \, {\text{i}}}{ \left( q^2 +{\text{i}}\epsilon \right)^2 } \int {\mathrm{d}}^4 \! u \; M^{ \left[ \rho\sigma \right] } \left( u
\right) \frac{ q_\sigma \,\left( - \kappa\, u_\rho \right) }{\left( 1 - 2 \, \kappa \, X + \kappa^2 \, X^2 M^2 \right)^2 }
\end{eqnarray}
where we have used the relation
\begin{eqnarray}
\label{h_partial}
 \partial_\rho^u \;\mathbf{h}_n^\nu(u,q) = q_\rho \,\mathbf{h}_{n-1}^{\nu+1}(u,q) - u_\rho \,\mathbf{h}_{n-2}^{\nu+1}(u,q)\;.
\end{eqnarray}
Again, all sub-leading twists, beginning with twist--5, vanish for the same reason as before. Therefore, also in the case of
$M \left( \kappa x , - \kappa x \right) $ the Fourier transformation acts as projection onto leading twist.

To complete the discussion we give the result for the operator $O^{\text{tw2}} \left( \kappa x , - \kappa x \right)$:
\begin{eqnarray}
\label{O_fourier}
 \int \frac{{\mathrm{d}}^4 \! x}{2\pi^2} \; {\text{e}}^{{\text{i}} qx} \frac{1}{ x^2 - {\text{i}}\epsilon} \; O^{\text{tw2}}\left( \kappa x,
- \kappa x \right)
&=&
 \sum_{n=0}^\infty \kappa^n \int {\mathrm{d}}^4 \! u \; O^\rho\!\left( u \right) \; \partial_\rho^u \; \mathbf{h}_{n+1}^1(u,q)
\\
\label{O_fourier_b}
&=&
 \sum_{n=0}^\infty \kappa^n \int {\mathrm{d}}^4 \! u \; O^\rho\!\left( u \right) \; \left( q_\rho \, \mathbf{h}_{n}^2(u,q) - u_\rho \,
\mathbf{h}_{n-1}^2(u,q) \right)
\\
\nonumber
&=&
 \frac{4}{ \left( q^2+{\text{i}}\epsilon \right)^2 } \int {\mathrm{d}}^4 \! u \; O^\rho\!\left( u \right) \frac{ q_\rho - \kappa \, u_\rho
}{\left( 1 - 2 \, \kappa \, X + \kappa^2 \, X^2 M^2 \right)^2 }\;.
\end{eqnarray}
Comparing equations (\ref{M_fourier_b}) and (\ref{O_fourier_b}) with equations (\ref{M3n}) and (\ref{O2n}) one again observes the same
structure in terms of Gegenbauer polynomials (Remind, that $\mathbf{h}_{n}^\nu(u,q)$ contains an additional factor of $(2/q^2)^{n+\nu}$
relative to ${h}_{n}^\nu(u,q)$).

As a comprising remark we conclude that the light-cone limit as well as the Fourier transformation with the light-cone singularity
$1 / \left( x^2 - {\text{i}} \epsilon \right)$ acts as a leading twist projection for scalar operators .
\subsection{ The operator $ O\left( \kappa x , - \kappa x \right) $ multiplied by the leading light-cone singularity $ 1 / ( x^2 - {\text{i}}\epsilon )^2 $ }
The scalar operator $ O\left( \kappa x , - \kappa x \right) $ appears in the Compton amplitude with the leading leading light-cone
singularity $ 1 / \left( x^2 - {\text{i}} \epsilon \right)^2 $. As a first step we provide the Fourier transformation of the complete
operator,
\begin{eqnarray}
\label{un_decomp_O}
 \int \frac{{\mathrm{d}}^4 \! x}{2\pi^2} \; {\text{e}}^{{\text{i}} qx} \frac{1}{\left( x^2 - {\text{i}}\epsilon \right)^2} \; O\left( \kappa
x, - \kappa x \right)
&=&
 \frac{1}{2} \sum_{n=0}^\infty \kappa^n \int {\mathrm{d}}^4 \! u \; O^\rho\!\left( u \right) \; \partial_\rho^u \; \mathbf{h}_{n+1}^0(u,q)
\\
\nonumber
&=&
 \frac{1}{q^2+{\text{i}}\epsilon} \int {\mathrm{d}}^4 \! u \; O^\rho \! \left( u \right) \frac{ q_\rho - \kappa \, u_\rho }{1 - 2\,\kappa
\,X + \kappa^2 \, X^2 M^2 }\;.
\end{eqnarray}
The technique used in the preceding section can be directly applied to the present case. One only needs the Fourier transform of
$ (ux)^n / \left( x^2 - {\text{i}} \epsilon \right)^2 $ followed by the differentiations $ \left( - {\text{i}} \, u \partial_q \right)^n $
which gives
\begin{equation}
\label{formel_wichtig_b}
 \int \frac{{\mathrm{d}}^4 \! x}{2\pi^2} \; {\text{e}}^{{\text{i}} qx} \frac{ \left( ux \right)^n }{ \left( x^2 - {\text{i}}\epsilon
\right)^2 } = - \frac{1}{2} \; {\text{i}}^{n+1} \; n! \;\; \mathbf{h}^0_n\left( u,q \right)\,,\quad n > 0\,.
\end{equation}
Equipped with this formula one gets the Fourier transform of
$ O^{\text{tw2}}\left( \kappa x, -\kappa x \right)/\left( x^2 -{\text{i}} \epsilon \right)^2$ as follows:
\begin{eqnarray}
\nonumber
 \int \frac{{\mathrm{d}}^4 \! x}{2\pi^2} \; {\text{e}}^{{\text{i}} qx} \frac{1}{\left( x^2 - {\text{i}}\epsilon \right)^2} \;
O^{\text{tw2}}\left( \kappa x, - \kappa x \right)
&=&
 \frac{1}{2} \sum_{n=0}^\infty \kappa^n \int {\mathrm{d}}^4 \! u \; O^\rho\!\left( u \right) \; \partial_\rho^u \; H_{n+1}\left( u^2,\Box_u
\right) \;\, \mathbf{h}_{n+1}^0(u,q)
\\
\label{O_fourier_c}
&=&
 \sum_{n=0}^\infty \frac{\kappa^n}{4\left( n+1 \right) } \int {\mathrm{d}}^4 \! u \; O^\rho\!\left( u \right) \; q^2 \,\partial_\rho^u \;
\mathbf{h}_{n+1}^1(u,q)
\\
\nonumber
&=&
 - \frac{1}{4\,\kappa} \int {\mathrm{d}}^4 \! u \; O^\rho\!\left( u \right) \; \partial_\rho^u \; \left( L_+ + \frac{ L_- }{ \sqrt{1-M^2} }
\right)\;.
\end{eqnarray}
Comparing equations (\ref{O_fourier}) and (\ref{O_fourier_c}) one realizes that the difference between the two expressions only consists
in an additional factor of $q^2 / \left( 4\left( n+1 \right) \right)$ for the latter one. This factor occurs since
$\Box_u \, \mathbf{h}^0_{n+1}(u,q)=-2 \, \mathbf{h}^1_{n-1}(u,q)$ and because of the fact that the following recurrence relation holds due
to a corresponding one for the Gegenbauer polynomials (see Ref.~\cite{PBM}, Appendix II.11)
\begin{equation}
\label{recurrence}
 2 \left( n + \nu \right) \, \mathbf{h}^\nu_n(u,q) = q^2 \, \mathbf{h}^{\nu+1}_n(u,q) - u^2 \, \mathbf{h}^{\nu+1}_{n-2}(u,q) \; .
\end{equation}
Summing up to a non-local operator is carried out by first substituting the factor $1/(n+1)$ by the integral
$\int_0^1 {\mathrm{d}}\tau \, \tau^n$ followed by a summation according to formula (\ref{summe}). As a last step the $\tau$-integration is
performed leading to a representation in terms of the functions $L_+$ and $L_-$ defined as
\begin{eqnarray*}
 L_\pm &=& \ln \left( 1 - \kappa \, X \bigl( 1 + \sqrt{1-M^2} \bigr) \right) \pm \ln \left( 1 - \kappa \, X \bigl( 1 - \sqrt{1-M^2} \bigr)
\right)\;.
\end{eqnarray*}
To prove the complete twist decomposition for the Fourier transformation of
$ O\left( \kappa x, - \kappa x \right) / ( x^2 - {\text{i}} \epsilon )^2 $ we straightforwardly calculate the twist--4 part by
\begin{eqnarray}
\label{O_fourier_d}
 \frac{1}{4} \int \frac{{\mathrm{d}}^4 \! x}{2\pi^2} \; {\text{e}}^{{\text{i}} qx} \frac{x^2}{\left( x^2 - {\text{i}}\epsilon \right)^2} \;
\int_0^1 {\mathrm{d}}\tau \; \tau \, O^{\text{tw4}}\left( \kappa \tau x, - \kappa \tau x \right)
&=&
 - \sum_{n=0}^\infty \frac{\kappa^n}{4\left( n+1 \right)} \int {\mathrm{d}}^4 \! u \; O^\rho\!\left( u \right) \; \partial_\rho^u \; u^2 \;
\mathbf{h}_{n-1}^1(u,q)
\\
\nonumber
&=&
 -\frac{1}{4\, \kappa} \int {\mathrm{d}}^4 \! u \; O^\rho\!\left( u \right) \; \partial_\rho^u \left( L_+ - \frac{ L_- }{ \sqrt{1-M^2} }
\right)\;,
\end{eqnarray}
where we have again used the integral representation of $1/(n+1)$, formula (\ref{formel_wichtig}) and the harmonicity of
$\mathbf{h}^1_{n-1}(u,q)$ with respect to $u$. Again, one observes the same structure in terms of Gegenbauer polynomials of the local
operators in $x-$ and $q-$space when the $q-$differentiation in (\ref{O_fourier_d}) is performed and compared with Eq.~(\ref{O4n}).

As expected, adding formulae (\ref{O_fourier_c}) and (\ref{O_fourier_d}) and using relation (\ref{recurrence}) one immediately recovers
the undecomposed operator (\ref{un_decomp_O}). Therefore, in terms of non-local operators the complete twist decomposition reads
\begin{equation*}
 \int \frac{{\mathrm{d}}^4 \! x}{2\pi^2} \; {\text{e}}^{{\text{i}} qx} \frac{1}{\left( x^2 - {\text{i}}\epsilon \right)^2} \; O\left( \kappa
x, - \kappa x \right) = \int \frac{{\mathrm{d}}^4 \! x}{2\pi^2} \; {\text{e}}^{{\text{i}} qx} \frac{1}{\left( x^2 - {\text{i}}\epsilon
\right)^2} \left( O^{\text{tw2}}\left( \kappa x, - \kappa x \right) + \frac{x^2}{4} \int_0^1 {\mathrm{d}}\tau \; \tau \,
O^{\text{tw4}}\left( \kappa \tau x, - \kappa \tau x \right) \right) \; .
\end{equation*}
Again the above results (\ref{O_fourier_c}) and (\ref{O_fourier_d}) can be obtained by calculating directly the Fourier transform of
(\ref{O4n}), thereby using the relation
\begin{equation*}
 \int \frac{{\mathrm{d}}^4 \! x}{2\pi^2} \; {\text{e}}^{{\text{i}} qx} \frac{1}{ \left( x^2 - {\text{i}}\epsilon \right)^2 } \; 
 h^\nu_n \left( u,x \right) = -
{\text{i}}^{n+1} \frac{ \left( n+\nu-2 \right)! }{ 4 \left( \nu-1 \right)! } \; \Bigl( q^2 \, \mathbf{h}^1_n \left( u,q \right)
 + 2 \, \left( \nu-1 \right) \, \mathbf{h}^0_n \left( u,q \right) \Bigr) \; .
\end{equation*}

\section{Conclusions}
In this Letter we have shown by explicit calculation that the infinite twist decomposition of the scalar off-cone operators
$ N \left( \kappa x , - \kappa x \right) $, $ O \left( \kappa x , - \kappa x \right) $ and $ M \left( \kappa x , - \kappa x \right) $ in
$x-$space as given in Ref.~\cite{GLR01} becomes finite after Fourier transformation with the corresponding singular functions appearing in
the expansion (\ref{str_wick}) of the product of hadronic currents near the light-cone. This twist cut-off depends on the order of the
corresponding light-cone singularity. The leading light-cone singularity $ 1 / \left( x^2 - {\text{i}} \epsilon \right)^2 $ projects out the
twist--2 and twist--4 content of the $O$-operator, while the sub-leading singularity $ 1 / \left( x^2 - {\text{i}} \epsilon \right) $ acts
as a projection onto leading twist for all considered operators. In addition, as a general result we found that all local off-cone scalar
operators of definite twist in $q-$space have exactly the same structure in terms of Gegenbauer polynomials as the corresponding ones in
$x-$space.

These observations are of immediate phenomenological interest. They show that in physical processes only a finite number of twists
contribute to target-mass corrections and that these contributions, in principle, can be read off from the off-cone twist decomposition in
$x-$space.

Of course, the scalar operators treated here are not the most important ones. Therefore, a corresponding twist cut-off may be assumed (and
will be proven) also for the non-local (axial) vector operators $O_\mu^{(5)} \left( \kappa x , - \kappa x \right) $ and the skew tensor
operator $M_{[\mu\nu]}(\kappa x, -\kappa x)$ which contribute to the leading and sub-leading light-cone singularity, respectively. In
addition, analogous considerations for the related vector operator $x^\nu M_{[\mu\nu]}(\kappa x, -\kappa x)$ are of interest, e.g., in the
case of Drell-Yan processes. These considerations are postponed to a subsequent article.

\acknowledgments
\noindent
The authors are grateful to M. Lazar, D. Robaschik and J. Bl\"umlein for various useful discussions. In addition, J. Eilers gratefully
acknowledges the Graduate College "Quantum field theory" at Center for Theoretical Studies of Leipzig University for financial support.

\end{document}